\documentstyle[11pt,aas2pp4]{article}	
      
\tightenlines

\def	\Angstrom {\,{\rm\AA}}
\def	\AU	{\,{\rm AU}}
\def	\beq	{\begin{equation}}
\def	\cm	{\,{\rm cm}}

\def	\eeq	{\end{equation}}

\def	\g	{\,{\rm g}}
\def	\gtsim	{\gtrsim}					  
\def	\H	{{\rm H}}
\def	\HH	{{\rm H}_2}
\def	\K	{\,{\rm K}}
\def	\kms	{\,{\rm km~s}^{-1}}
\def	\kpc	{\,{\rm kpc}}

\def	\ltsim	{\lesssim}					  
\def	\Msol	{M_{\odot}}
\def	\micron	{\,\mu{\rm m}}

\def	\yr	{\,{\rm yr}}

\lefthead{Draine}
\righthead{Lensing of Stars by Gas Clouds}

\begin{document}
\begin{center}
\noindent{\bf POPe-762}\hfill Submitted to {\it The Astrophysical Journal (Letters)}
\end{center}
\bigskip\bigskip
\title{Lensing of Stars by Spherical Gas Clouds}

\author{B.T. Draine}
\affil{Princeton University Observatory, Peyton Hall, Princeton,
NJ 08544; draine@astro.princeton.edu}

\begin{abstract}
If the Galaxy contains $\sim\!10^{11}\Msol$ in 
cold gas clouds of $\sim$Jovian mass and $\sim$AU size,
these clouds will act as converging lenses for optical light,
magnifying background stars at a detectable rate.
The resulting light curves can
resemble those due to gravitational lensing by a point mass, raising the
possibility that some of the events attributed to gravitational microlensing
might in fact be due to ``gaseous lensing''.
During a lensing event, 
the lens would impose
narrow infrared and far-red $\HH$ absorption lines 
on the stellar spectrum.
Existing programs to observe gravitational microlensing, supplemented
by spectroscopy, can therefore
be used to either detect such events or place limits on the number
of such gas clouds present in the Galaxy.
\end{abstract}


\section{Introduction}

A number of authors have proposed that the Galaxy could contain a 
hitherto-unrecognized population
of small, cold, dense self-gravitating gas clouds, in numbers sufficient 
to contribute an appreciable fraction of the gravitational mass of the
Galaxy (Pfenniger, Combes, \& Martinet 1994; Gerhardt \& Silk 1996;
Combes \& Pfenniger 1997).

Walker \& Wardle (1998; hereafter WW98) 
pointed out that if such clouds
existed in the Galactic halo, each would have an ionized
envelope which could explain the ``Extreme
Scattering Events'', or ``ESEs'', (Fiedler et al. 1987)
during which extragalactic point radio sources occasionally
undergo substantial frequency-dependent 
amplification and deamplification, apparently
due to refraction by a plasma ``lens'' moving across the line-of-sight.
WW98 proposed that the observed frequency of such ESEs
could be explained if there was 
a population of $\sim\!10^{14}$
cold self-gravitating gas clouds, with mass $M\approx 10^{-3}\Msol$,
and radius $R\approx 3\AU$.

Gerhardt \& Silk (1996) and WW98 noted that
if the clouds were opaque, their presence would have been revealed by
existing stellar monitoring programs studying gravitational
``microlensing'' (Paczynski 1986; see the review by Paczynski 1996, and
references therein),
as these experiments would have detected occultation events of duration
$\sim\! R/200\kms\approx 40 {\rm \,days}$.
Such occultation events have not been reported.
However, the hypothesized clouds could be essentially transparent at
optical wavelengths: they could have
formed from primordial gas, or, if formed from gas containing metals,
the grains could have sedimented to form a small core.

In this {\it Letter} we point out that even transparent clouds would
have lensing effects which would be detectable by the stellar
monitoring studies currently underway to study gravitational lensing by compact
objects in our Galaxy.
Existing data can thus test the hypothesis that cold gas clouds
contribute an appreciable
fraction of the mass of the Galaxy.

\section{Density Profile\label{sec:rho(r)}}

We consider nonrotating polytropic models for self-gravitating
H$_2$-He gas clouds of radius $R$.
The polytropic index $n$ ($T\propto\rho^{1/n}$) 
is assumed to be in the range $1.5<n<5$;
for $n<1.5$ the cloud would be convectively unstable, while for
$n\ge5$ the central density is infinite.
We do not expect $T(r)$ and
$\rho(r)$ to be accurately described by a polytropic model,
but a slight rise in temperature toward the interior may be reasonable
since the interior will be heated by high energy cosmic rays,
with the few cooling lines [e.g., H$_2$0-0S(0) 28.28$\micron$, or
HD 0-0R(1) 112$\micron$]
very optically thick.
Furthermore, these polytropes have $T(R)=0$, so the density structure
near the surface is unphysical.
Table \ref{tab:polytropes} gives various properties 
for H$_2$-He polytropes, including the half-mass radius $r_h$, 
$T_c\equiv T(0)$, $T_h\equiv T(r_h)$, and $\rho_c\equiv\rho(0)$ and
$\rho_h\equiv\rho(r_h)$ relative to the
mean density
$\langle\rho\rangle \equiv {3M/ 4\pi R^3}$.

\placetable{tab:polytropes}

\section{Gaseous Lensing\label{sec:lensing}}

For $\rho\ltsim10^{-2}\g\cm^{-3}$, the refractive index $m$ is
\beq
m(\lambda) = 1 + \alpha(\lambda)\rho ~~~;
\eeq
$\alpha(4400\Angstrom)= 1.243 \cm^3\g^{-1}$ and
$\alpha(6700\Angstrom)= 1.214 \cm^3\g^{-1}$
for H$_2$/He gas with 24\% He by mass 
(AIP Handbook 1972).

\begin{figure}
\epsscale{1.00}	
\plotfiddle{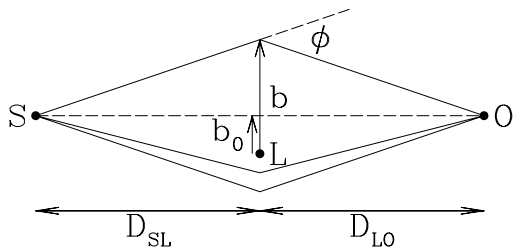}{3.0cm}{-90}{25}{25}{-100}{140}
\figcaption[f1.eps]{
	\label{fig:geom}
	Geometry of lensing. S is the source, the center of the lens is
	at L, and O is the observer.
	}
\end{figure}

For small deflections, 
a light ray with impact parameter $b$ will be deflected through
an angle
\beq
\phi(b) = -2\alpha b \int_b^\infty {dr \over (r^2-b^2)^{1/2}}
{d\rho\over dr}  ~~~.
\label{eq:phi}
\eeq
There is a small additional gravitational deflection
(Henriksen \& Widrow 1995), but this is negligible compared to
gaseous refraction.

Let $D_{\rm SL}$ and and $D_{\rm LO}$ be 
the distance from source to lens, and
from lens to observer.
If $b_0$ is the distance of the lens center from the 
straight line from source
to observer (see Fig.\ \ref{fig:geom}), 
then the apparent distance $b$ of the image from the
lens is given by the lensing equation
\beq
b - b_0 = D\phi(b) ~~~,~~~
D \equiv {D_{\rm SL}D_{\rm LO}\over D_{\rm SL}+D_{\rm LO}} ~~~.
\label{eq:lensing}
\eeq

For a point source,
the image magnification is given by
\beq
M(b) = {|b|\over b_0}{1 \over 1-D\phi^\prime(b)}  ~~~,
\label{eq:mag}
\eeq
\beq
\phi^\prime(b) \equiv {d\phi(b)\over db} =  -2\alpha\int_0^\infty dz 
\left[
{b^2\over r^2}{d^2\rho\over dr^2} + {z^2\over r^3}{d\rho\over dr}
\right] ~~~.
\eeq
where $r^2=b^2+z^2$.
For a given $b_0$ 
there will be an odd number $N(b_0)$ of solutions $b_i(b_0)$, $i=1,...,N$.
The total 
amplification $A(b_0)=\sum_{i=1}^N M(b_i)$.
The ``trajectory'' of the lens relative to the source is characterized by
a ``source impact parameter'' $p$ and a displacement $x$ along the
trajectory; for any $x$ we have $b_0=(p^2+x^2)^{1/2}$, and the
``light curve'' is just $A(b_0)$ {\it vs.} $x$.

We define a ``strength'' parameter
\beq
S \equiv {\alpha\langle\rho\rangle D \over R} = 
3.55\left({M\over10^{-3}\Msol}\right)\left({\AU\over R}\right)^4
\left({D\over 10\kpc}\right) ~~~.
\label{eq:Sdef}
\eeq
When $S$ is of order unity, large
amplifications (and deamplifications) are possible even
for $b_0/R$ of order unity, whereas
if $S\ll 1$, appreciable amplification only occurs for $b_0\ll R$.

\section{Sample Light Curves}

For each polytropic index 
we define the critical strength $S_c(n)$ to be the value of
$S$ such that for $b_0=0$ 
there are 3 images for $S>S_c$ and one image for $S<S_c$.
Values of $S_c$ are given in Table \ref{tab:polytropes}.

For each case with $S>S_c$ we define the critical source
impact parameter $b_{0c}(S)$ such that there is one image for $b_0>b_{0c}$,
and 3 images for $b_0<b_{0c}$.
For $S>S_c$ a light curve for which $p<b_{0c}$ will have
``cusps'' at the points where $b_0=b_{0c}$, as two images either appear
(for $|x|$ decreasing) or merge and disappear (for
$|x|$ increasing).

\begin{figure}
\epsscale{1.00} 
\plotone{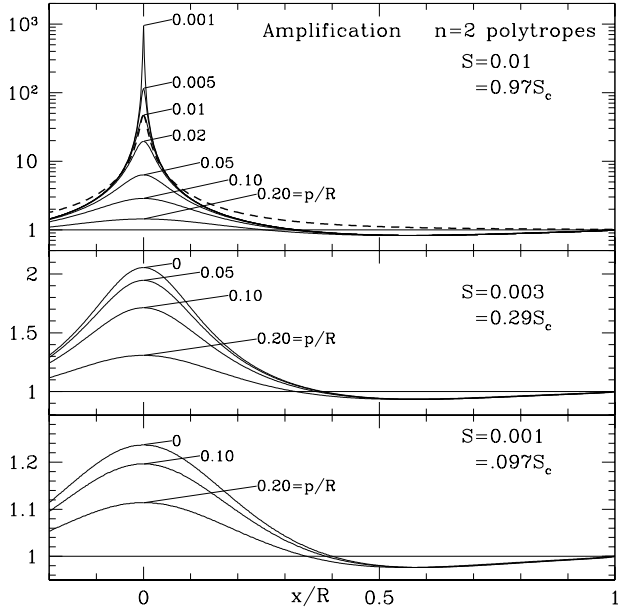}
\figcaption[f2.eps]{
	\label{fig:n=2}
	Light curves for transparent polytropes of index
	$n=2$ and lensing strength $S=0.001$, 0.003, and 0.01.
	Also shown is an ideal ``gravitational lensing'' light curve
	(broken line) fitted to the light curve for $p=0.01R$.
	For these three cases, with $S<S_c$, 
	the gaseous lens 
	light curves resemble gravitational lensing light curves 
	when the amplification is large,
	but show systematic differences at low amplification; in particular,
	every gaseous lens has a range of $b_0$ where it deamplifies.
	}
\end{figure}

\begin{figure}
\epsscale{1.00} 
\plotone{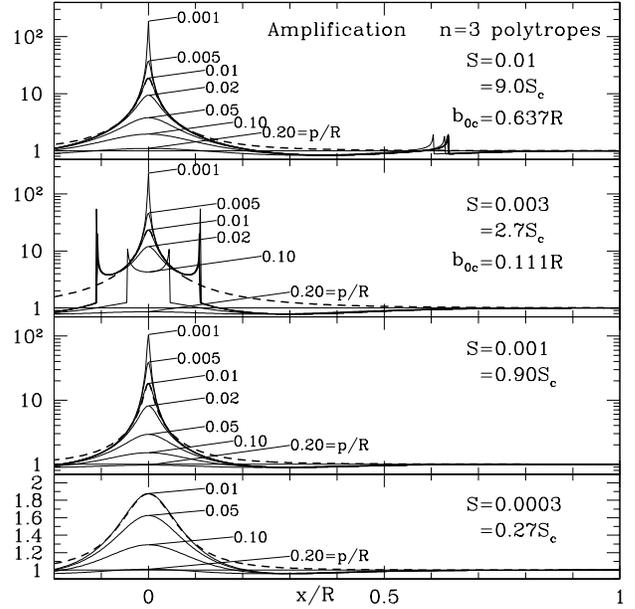}
\figcaption[f3.eps]{
	\label{fig:n=3}
	Same as Fig.\ \protect{\ref{fig:n=2}}  
	but for $n=3$.
	When $S>S_c$, light curves
	with $p<b_{0c}(S)$ contain conspicuous cusps (see text).
	}
\end{figure}

\begin{figure}
\epsscale{1.00} 
\plotone{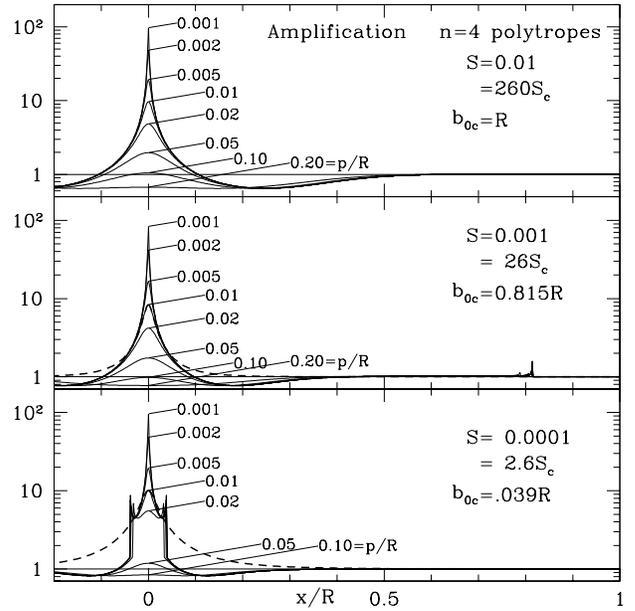}
\figcaption[f4.eps]{
	\label{fig:n=4}
	Same as Fig.\ \protect{\ref{fig:n=2}} 
	but for $n=4$.
	}
\end{figure}

Figures \ref{fig:n=2}--\ref{fig:n=4} 
show light curves for lenses with polytropic indices $n=2$, 3, and 4,
respectively.
The following general behavior is found:
(1) For $S\ltsim 0.3S_c$ the lensing is weak, with
peak amplification $A(0)\ltsim2$;
(2) For $0.7S_c \ltsim S < S_c$ the peak amplification can be very
large, even though only one image is formed;
(3) For $S_c< S\ltsim 5S_c$ the peak amplification is very large, and
light curves with $p<b_{0c}$ 
have conspicuous cusps where the amplification for a point
source is infinite;
(4) For $S\gtsim 10S_c$, the cusps become weak because image merging
only occurs for $b_0\approx R$, and the two merging images are faint.

In each figure we also show, for comparison, 
an ideal ``gravitational lensing'' light curve
fitted\footnote{
	The gravitational lensing light curve is fitted by requiring it
	to have the same peak amplification 
	$A_{max}$ and to have the same width at
	$A=A_{max}^{1/2}$.
	}
to the light curve for $p=0.01R$.
Note that for $S<S_c$ (no cusps present) or 
$S\gtsim10S_c$ ($b_{0c}\gtsim 0.7R$, so that the cusps are weak) 
there is considerable similarity between the gravitational
lensing and gaseous lensing light curves.

\section{Discussion\label{sec:discuss}}

For purposes of discussion, we adopt 
the $n=3$ polytropic model (with $\rho_c=54\langle\rho\rangle$)
as a guide.
For $S\gtsim0.5S_c$ we obtain amplifications $A(p)>1.5$ for $p\ltsim0.1R$
(see Fig.\ \ref{fig:n=3}).

\subsection{Lensing Event Rate}

Fiedler et al. (1994) report 9 ESEs in 594 source-years of monitoring.
Two events (0954+658 and 1749+096, with durations 
$\sim\!0.35\yr$) have radio light curves
suggestive of lensing by the ionized atmosphere of a spherical cloud,
implying a rate 
$\dot{P}_{\rm ESE}\approx2/594\approx3\!\times\!10^{-3}\yr^{-1}$ per source,
and a ``covering factor'' 
$f_{\rm ESE}\approx\dot{P}_{\rm ESE}\!\times\!0.35\yr\approx
1\!\times\!10^{-3}$.
If the plasma envelope extends to $\sim\!1.5R$, then
from the 0.35yr duration of the radio events we estimate 
$R\approx200\kms\!\times\!0.35\yr/2\approx7\AU$.

The neutral cloud covering factor would then be 
$f\approx f_{\rm ESE}/(1.5)^2\approx 5\!\times\!10^{-4}$,
and the rate per source of optical lensing events with $p<0.1R$ is
\beq
\dot{P}_{\rm OL}\approx 
(0.1R / 1.5R)
\dot{P}_{\rm ESE}\approx 2\!\times\!10^{-4} \yr^{-1} ~~~.
\label{eq:Pdot_OL}
\eeq
The optical light curve would have a characteristic time scale
$\sim\! 0.2R/(200\kms)\approx 10$ days,
short compared to the observed lensing events
toward the LMC (Alcock et al. 1997), but within the detectable range:
with an effective exposure of $\sim\!2\!\times\!10^6$ source-years 
the absence of $\sim\!10$ day events exceeding the $A>1.75$
MACHO threshold implies an upper limit of
$\sim\!1.5\!\times\!10^{-6}\yr^{-1}$,
100 times smaller than $\dot{P}_{\rm OL}$ from eq.\ (\ref{eq:Pdot_OL})!
We can therefore exclude the possibility that the typical cloud
can produce amplifications $A\gtsim 2$:
the gas clouds associated with ESEs must have $S\ltsim 0.3S_c$ for
$D\approx10\kpc$.
Thus
\beq
M \ltsim 8.5\!\times\!10^{-5}\Msol (R/\AU)^4 (10\kpc/D) S_c ~~~.
\label{eq:limScrit}
\eeq
\subsection{Other Limitations}
\begin{figure}
\epsscale{1.00}
\plotone{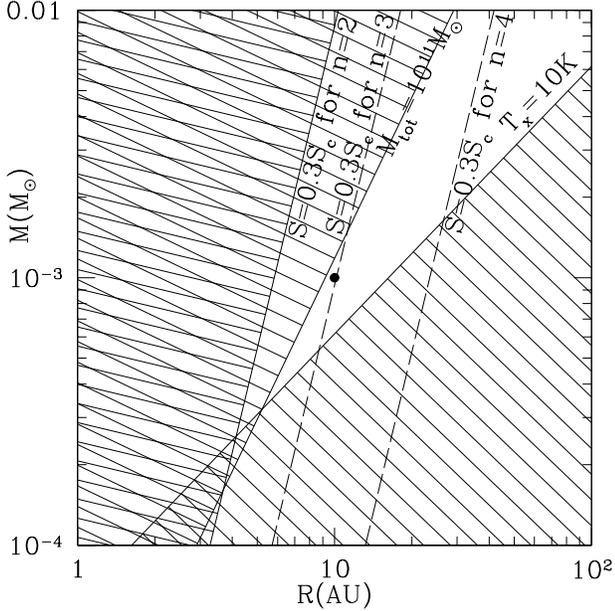}
\figcaption[f5.eps]{
	\label{fig:forbid}
	Allowed values of $M$ and $R$, for gas clouds at $D\approx10\kpc$.
	Below the line ``$T_x = 10\K$'' the surface layers of the
	polytrope would be unbound for $T=10\K$.
	Above the line ``$S=0.3S_c$'' the cloud
	would be capable of
	amplification $A\gtsim2$ and hence this region is ruled out by
	searches for optical lensing toward the LMC.
	Above the line $M_{tot}=10^{11}\Msol$ clouds with a covering
	fraction $f=5\!\times\!10^{-4}$ would contribute 
	more than $10^{11}\Msol$.
	}
\end{figure}

If the clouds are at a typical distance
of $10\kpc$ and contribute a covering fraction $f\approx5\!\times\!10^{-4}$
(see above), the total mass in clouds should not exceed $\sim\!10^{11}\Msol$:
\beq
M \ltsim 1\!\times\!10^{-5}\Msol (R/\AU)^2 ~~~.
\label{eq:limMtot}
\eeq
A third requirement is that the gravitational binding energy of
gas particles near the surface exceed the thermal kinetic energy
per particle for plausible surface temperature 
$T\approx10\K$:
\beq
M \gtsim 6\!\times\!10^{-5}\Msol (R/\AU) ~~~.
\label{eq:limTx}
\eeq
These 3 conditions are plotted in Fig.\ \ref{fig:forbid}.
We see that the parameters favored by WW98 -- $M=10^{-3}\Msol$, $R=3\AU$ -- 
are ruled out by both eqs.\ (\ref{eq:limScrit}) and (\ref{eq:limMtot}).
However, an allowed region does remain,
including $M\approx10^{-3}\Msol$, $R\approx10\AU$.

\subsection{Demagnification}

For $p > b_{0c}$ ({\it i.e.}, no cusps),
light curves for spherical gaseous lenses 
bear considerable 
similarity to light curves for gravitational lensing.
As a consequence, it would not be trivial to distinguish gravitational
lensing from gaseous lensing purely on the basis of optical light
curves, particularly since departures from ideal gravitational lensing
of point sources by point masses are anticipated due to blending with other
stars, lensing by stars with planetary or stellar companions, and
sources of finite angular extent.
The most conspicuous difference in the light curves is the fact that
gaseous lensing always has a range of $b_0$ values for which
$A<1$, unlike gravitational lensing.
Accurate photometry could either detect such deamplification, or place
a limit on it.
Note that for lenses with $S\ll1$ the deamplification is very slight:
for example, for $n=3$ and $S<S_c$, $A>0.88$.
Thus accurate photometry would be required to either detect such
deamplification or obtain a useful limit.

\subsection{Spectroscopy}
The cold molecular gas will be nearly transparent at infrared and optical
frequencies, but could be detected through H$_2$ absorption lines.
The characteristic H$_2$ column density is
\beq
N(\HH)\approx {M/R^2\over 2.64m_\H}\approx
2\!\times\!10^{25}\!\left(\!{M\over10^{-3}\Msol}\!\right)\!
\left(\!{10\AU\over R}\!\right)^2\!\cm^{-2}.
\label{eq:NH2typ}
\eeq
The $\HH$ will be mainly in the $(v,J)=(0,0)$ and $(0,1)$ levels.
Quadrupole vibrational absorption lines out of these two levels
are listed in Table \ref{tab:H2lines}, with line-center absorption
cross sections $\sigma_0$ computed neglecting pressure broadening and
assuming only thermal broadening at $T=20\K$.
For the estimated column densities in eq.\ (\ref{eq:NH2typ}) we see
that the 1-0 transitions would have central optical depths
$\tau\approx 10^3$.
While the overtone transitions are weaker, the increased stellar
brightness near $\sim\!8300\Angstrom$ plus sensitive CCD detectors
may make the 3-0 transitions best to use.
Detection of these absorption features would both confirm the
gaseous nature of the lens and determine its radial velocity, 
expected to be $\sim\! 200 \kms$ if the lens belongs to the
halo as proposed by WW98.

\placetable{tab:H2lines}

\subsection{Chromaticity}
The optical dispersion results in a slightly larger value of $S$ in the blue.
However, noting the similarity in light curves for $S=0.01$ and $S=0.001$
for $n=3$, it is clear that changing $S$ by 0.8\% will be expected to produce
only a very slight increase in amplification (``blueing'').
If, however, caustics are present in the light curve, the caustics will
occur at different times for different colors, which would be
detectable with multicolor photometry with
good time resolution.

The increased amplification at shorter wavelengths is counteracted
by Rayleigh scattering, which will
redden light passing through the cloud (WW98).
The Rayleigh scattering cross section is
$\sim\! 8.4\!\times\!10^{-29}(\micron/\lambda)^4\cm^2$, resulting in
reddening of a magnified star by 
$E(B-V)\approx 1.3\!\times\!10^{-27}N(\HH)\cm^2$, giving $E(B-V)\approx 0.03$
for $N(\HH)$ from eq.\ (\ref{eq:NH2typ}).

If the lensed star happened to be bright in the ultraviolet, the gaseous
lens would completely block the ultraviolet radiation shortward of
$\sim\!1200\Angstrom$ through the damping wings of the Lyman and Werner
band transition of $\HH$, which overlap to form an opaque continuum
shortward of $\sim\!1110\Angstrom$ for $N(\HH)\gtsim 10^{21}\cm^{-2}$
(Draine \& Bertoldi 1996).
Since most target stars are far too faint to be detected
in the vacuum ultraviolet, however, it seems unlikely that this absorption
could be observed.

\section{CONCLUSIONS}

It is not obvious how 
cold self-gravitating gas clouds with the properties suggested by
WW98 might have formed, or whether such clouds would
be stable for $\sim\!10^{10}\yr$.
However, if they do exist, WW98 show that they could
solve two longstanding problems:
(1) their ionized envelopes could account for some of the ``Extreme
Scattering Events''; and
(2) they could contain the ``missing'' baryons in the Galaxy.
It is notable that these same clouds could 
ameliorate a third problem: the fact that microlensing searches
detect a larger number of amplification events toward the LMC than 
expected for lensing by stars and stellar remnants.
Some of these events could be due to gaseous lensing.
Indeed, lensing by the hypothesized clouds would be so frequent that existing
programs to observe gravitational microlensing can already place
strong limits on the cloud parameters (see Fig.\ \ref{fig:forbid}),
but clouds with $M\approx10^{-3}\Msol$ and $R\approx10\AU$
are still allowed.

With a predicted 
lensing rate $\dot{P}_{\rm OL}\approx 2\!\times\!10^{-4}\yr^{-1}$,
the typical lens must be weak, with $S\ltsim 0.3S_c$.
The distribution of cloud properties and distances
could produce occasional strong gaseous lensing events
with $S\gtsim 1$, perhaps
accounting for some of the events attributed to gravitational
microlensing.

For non-caustic lensing,
the dispersive effects of the gas would produce slightly larger
amplification in the blue, but this is counteracted by
Rayleigh scattering by the $\HH$.
A number of quadrupole lines of
$\HH$ would be detectable in absorption during the lensing event; this would
be the most unambiguous signature of ``gaseous lensing''.

Existing programs to observe gravitational lensing, supplemented by
spectroscopy during lensing events,
can therefore be used to either detect gaseous lensing events or
place limits on the number of $\sim\!10^{-3}\Msol$ $\HH$ clouds
in the Galaxy.

\acknowledgements

I am grateful to Bohdan Paczynski for helpful comments, and
to
Robert Lupton for the availability of the SM package.
This work was supported in part by
NSF grant AST-9619429.



\begin{center}
\begin{deluxetable}{ c c c c c c c }
\tablecolumns{7}
\tablewidth{0pc}
\tablecaption{Polytropic Models\label{tab:polytropes}}
\tablehead{
\colhead{$n$}&
\colhead{$\rho_c/\langle\rho\rangle$}&
\colhead{$\rho_h/\langle\rho\rangle$}&
\colhead{$r_h/R$}&
\colhead{$T_c$\tablenotemark{a}}&
\colhead{$T_h$\tablenotemark{a}}&
\colhead{$S_c$}
	}
\startdata
1.5&	5.9907&	2.3228&	0.52118&	13.19&	\ 7.52&	.026\\
2&	11.403&	3.5602&	0.43921&	14.74&	\ 7.81&	.0103\\
2.5&	23.407&	5.9809&	0.36004&	17.14&	\ 8.38&	.0037\\
3&	54.183&	11.416&	0.28331&	20.93&	\ 9.29&	.00111\\
3.5&	152.88&	26.539&	0.20879&	27.45&	10.76&	.00028\\
4&	622.41&	88.183&	0.13650&	40.80&	13.32&	$3.9\!\times\!10^{-5}$\\
4.5&	6189.5&	703.43&	0.06671&	81.60&	19.31&	$2\!\times\!10^{-6}$\\
\enddata
\tablenotetext{a}{For $M=10^{-3}\Msol$ and $R=10\AU$.}
\end{deluxetable}
\end{center}

\begin{center}
\begin{deluxetable}{ c c c c c c }
\tablecolumns{6}
\tablewidth{0pc}
\tablecaption{H$_2$ Absorption Lines\label{tab:H2lines}}
\tablehead{
\colhead{line}&
\colhead{$\lambda(\micron)$}&
\colhead{$\sigma_0$($\cm^2$)}&
\colhead{line}&
\colhead{$\lambda(\micron)$}&
\colhead{$\sigma_0$($\cm^2$)}
	}
\startdata
1-0 Q(1)&2.4066&$3.3\!\times\!10^{-24}$&3-0 Q(1)&0.8500&$8.8\!\times\!10^{-27}$\\
1-0 S(0)&2.2233&$7.7\!\times\!10^{-24}$&3-0 S(0)&0.8275&$3.4\!\times\!10^{-26}$\\
1-0 S(1)&2.1218&$4.3\!\times\!10^{-24}$&3-0 S(1)&0.8153&$2.5\!\times\!10^{-26}$\\
2-0 Q(1)&1.2383&$2.0\!\times\!10^{-25}$&4-0 Q(1)&0.6567&$5.6\!\times\!10^{-28}$\\
2-0 S(0)&1.1896&$5.9\!\times\!10^{-25}$&4-0 S(0)&0.6437&$2.8\!\times\!10^{-27}$\\
2-0 S(1)&1.1622&$3.9\!\times\!10^{-25}$&4-0 S(1)&0.6370&$2.4\!\times\!10^{-27}$\\
\enddata
\end{deluxetable}
\end{center}

\end{document}